\documentclass[10pt]{article} \usepackage{emulateapj}

\newcommand{\FeII}{Fe$\;${\small\rm II}\relax}

\newcommand{\OVI}{O$\;${\small\rm VI}\relax}
\newcommand{\CIV}{C$\;${\small\rm IV}\relax}
\newcommand{\htwo}{H$_2$}

\newcommand{\err}[2]{\ensuremath{^{+ #1}_{- #2}}}

\newcommand{\kms}{km~s$^{-1}$\relax}

\newcommand{\copernicus}{{\em Copernicus}\relax}

\newcommand{\fuse}{{\em FUSE}\relax}
\newcommand{\iue}{IUE}

\begin{document}

\title{\fuse\ Observations of Interstellar Gas Towards the LMC Star Sk~-67~05}

\author{S.D. Friedman\altaffilmark{1}, J.C. Howk\altaffilmark{1}, 
B-G Andersson\altaffilmark{1}, K.R. Sembach\altaffilmark{1},
T.B. Ake\altaffilmark{1}, K. Roth\altaffilmark{1}, D.J. Sahnow\altaffilmark{1},
B.D. Savage\altaffilmark{2}, D.G. York\altaffilmark{3},
G. Sonneborn\altaffilmark{4}, A. Vidal-Madjar\altaffilmark{5},
E. Wilkinson\altaffilmark{6}}

\altaffiltext{1}{Dept. of Physics \& Astronomy, 
	The Johns Hopkins University, Baltimore, MD 21218}

\altaffiltext{2}{Dept. of Astronomy, University of Wisconsin, 
        Madison, WI 53706}

\altaffiltext{3}{Dept. of Astronomy \& Astrophysics, University of Chicago, 
	Chicago IL 60637}

\altaffiltext{4}{Laboratory for Astronomy and Solar Physics,
NASA/GSFC, Code 681, Greenbelt, MD 20771}

\altaffiltext{5}{Institut d'Astrophysique de Paris, CNRS, 98 bis bld Arago, F-75014
Paris, France}

\altaffiltext{6}{Center for Astrophysics and Space Astronomy, University of
Colorado, CB 389, Boulder, CO 80309}

\begin{abstract} We report on measurements of interstellar \OVI, \htwo,
 \ion{P}{2}, \ion{Si}{2}, \ion{Ar}{1}, and \ion{Fe}{2} absorption along the line
 of sight to Sk~-67~05, a B0 Ia star in a diffuse \ion{H}{2} region in the
 western edge of the Large Magellanic Cloud (LMC).  We find $\log N(\mbox{\OVI})
 = 14.40 \pm 0.04$ in the Milky Way (MW) component and, using the \CIV\ column
 density from previous \iue\ observations, $N(\mbox{\CIV}) / N(\mbox{\OVI}) =
 1.00 \pm 0.16$, a value similar to other halo measurements made with \fuse.
 In the LMC component, $\log N(\mbox{\OVI}) = 13.89\pm0.05$, and $N(\mbox{\CIV})
 / N(\mbox{\OVI}) < 0.4$ ($3\sigma$), since only an upper limit on
 $N(\mbox{\CIV})$ is available.  Along this sightline the LMC is rich in
 molecular hydrogen, $\log N({\rm H}_2) = 19.50\pm0.08$; in the MW $\log
 N({\rm H}_2) = 14.95\pm0.08$.  A two-component fit for the excitation
 temperature of the molecular gas in the LMC gives $T_{01} = 59\pm5$ K for
 $J=0,1$ and $T_{ex}=800\pm330$ K for $J=3,4,5$.  For the MW, $T_{01} =
 99\err{30}{20}$ K; no excitation temperature could be determined for the higher
 rotational states.  The MW and LMC gas-phase [Fe/P] abundances are $\sim0.6$
 and $\sim0.7$ dex lower, respectively, than solar system abundances.  These
 values are similar to [Fe/Zn] measurements for the MW and LMC towards SN 1987A.

\end{abstract}



\section{Introduction}

The analysis of interstellar absorption lines provides fundamental information
about the content and physical conditions of the interstellar medium (ISM).
Absorption line spectroscopy can be used to study the ISM in our Galaxy, in
nearby systems such as the Magellanic Clouds, and in the intergalactic medium
out to the most distant QSOs.  These studies provide an opportunity to compare
elemental abundances and physical conditions in regions with differing chemical
histories.

As part of a general program to investigate the interstellar medium of the Milky
Way and nearby galaxies, we have used the {\em Far Ultraviolet Spectroscopic
Explorer} (\fuse) satellite (Moos et al.  2000) to observe the star Sk~-67~05
(HD~268605) in the Large Magellanic Cloud (LMC).  The first observations of
Galactic halo gas with \iue\ were along sightlines to stars in the LMC (Savage
\& de Boer 1979; 1981).  In this Letter we discuss the first \fuse\ observations
of an LMC star revealing \OVI\ absorption in both the Milky Way and LMC.  We
also discuss the measurements of \htwo, \ion{P}{2}, \ion{Si}{2}, \ion{Ar}{1},
and \ion{Fe}{2} absorption along this sightline.

\section{Observations and Data Processing}

Sk~-67~05 ($l$ = 278\fdg89, $b$ = -36\fdg32) is a B0 Ia star (Smith Neubig \&
Bruhweiler 1999) located near the western edge of the LMC.  It lies in a
diffuse \ion{H}{2} region (Chu et al.  1994) with relatively low diffuse X-ray
emission compared to regions closer to the center of the LMC (Snowden \& Petre
1994).  Ardeberg et al.  (1972) give \bv = -0.12, and for (\bv)$_{0}$ = -0.23
(Binney \& Merrifield 1998), we find E(\bv) = 0.11.  Using the Galactic
gas-to-dust correlation (Diplas \& Savage 1994) we infer $N$(\ion{H}{1}) $\sim
5.4 \times 10^{20} {\rm cm}^{-2}$ along this sightline.

This star was observed during the In-Orbit Checkout phase of the mission
(Sahnow et al.  2000) at various times between 1999 August 20 and 1999 October
19.  As these observations were part of tests designed primarily to align the
four optical channels in the instrument, the star was stepped across the $30
\arcsec \times 30 \arcsec$ aperture during the observations.  However, the data
were taken in time-tagged photon-address mode, so that corrections could be
made for this image motion.  The analysis here uses LiF1 (990 - 1080~\AA) data
only; the LiF2 channel has lower sensitivity and poorer flatfield
characteristics, and during this period no flatfield corrections were
available.  The SiC (905 - 1100~\AA) channels were generally not aligned.  The
instrument was still in its preflight focus configuration, and the spectral
resolution was $\lambda /\Delta \lambda \la 15,000$.  The wavelength scale was
established with a pre-flight dispersion solution.  However, because of the
stepping manner in which the data were obtained, additional zero point
adjustments were required.  Relative velocities are generally accurate to
$\sim$10 \kms over limited spectral ranges, and the absolute scale was set by
comparison with \iue\ spectra.  The data presented represent a total of
approximately 33 ksec of on-target exposure time for the LiF1A and 35 ksec for
the LiF1B spectral regions, approximately equally split between day and night.
Additional details of the data processing for this data set can be found
elsewhere (Massa et al.  2000).

This star exhibits variable stellar wind features.  However, as shown in Figure
3 of Massa et al.  (2000), the variablility is minimal at the wavelengths of
interest here, and does not affect our conclusions.

\section{Interstellar Absorption Features}

Figure 1 shows the composite spectrum from the LiF1 channel.  Several metal
lines that appear in the ISM of both the Milky Way and LMC are identified.  Most
other lines are due to \htwo\ absorption in the LMC.  The strong emission line
is terrestrial Ly$\beta \; \lambda1025.72$ airglow from the daytime exposures.

In this analysis we have used only the 1031.93 \AA\ member of the \OVI\ doublet
since the 1037.62 \AA\ line is blended with \ion{C}{2}$^*$ 1037.02 \AA\ and
several \htwo\ lines.  Figure 2 shows the \fuse\ \OVI\ $\lambda1031.93$
absorption line and a high-dispersion \iue\ spectrum of the \ion{C}{4}
$\lambda1550.77$.  The \htwo\ (6-0) P(3) line originating in the LMC appears at
+88 \kms, and must be modelled and removed to get an accurate estimate of the
\OVI\ column density.

We have analyzed the (2-0) to (9-0) \htwo\ Lyman bands, using the method
described in Shull et al.  (2000) to determine the equivalent widths,
$b$-values, and column densities for the Milky Way and LMC components of \htwo.
Since the typical spacing between the rotational-vibrational lines within each
Lyman band is very close to the spacing between LMC and the Milky Way
absorption, the measurement of the Milky Way \htwo\ depends on a careful
decomposition of blended lines.  Because of this difficulty, we used only the
(4-0) and (2-0) Lyman bands to measure the $J=0,1$ levels from the Milky Way
gas.

The results of the \htwo\ model, discussed below, give $\log$ N(\htwo) = 15.28
for the (6-0) P(3) in the LMC.  This was convolved with the instrumental
resolution of 25 \kms\ and divided out of the original spectrum to remove the
effects of the \htwo\ absorption.  The resulting \OVI\ profile is shown as a
light line in Fig.  2.

To calculate the \OVI\ absorption we have considered two possible continuum
placements, designated ``high'' and ``low,'' which are displayed in Fig.  2 as
long-dashed and dash-dotted lines, respectively.  The arrow at +180 \kms\
denotes the velocity we have adopted as separating the Milky Way and LMC
components of \OVI.  Table 1 gives \OVI\ equivalent widths derived using both
continua.

Figure 3 shows the spectra of several important metal lines and a molecular
hydrogen line along the sightline to Sk~-67~05.  For comparison, the \iue\
spectrum of \ion{Si}{2} $\lambda$1808.01 is shown.  The individual absorption
lines have separately been shifted in velocity to align the Milky Way components
with the corresponding feature in the \iue\ spectrum.  We established the
velocity scale for the \OVI\ region by shifting the \ion{Si}{2} $\lambda$1020.70
line to match the \ion{Si}{2} $\lambda$1808.01 \iue\ velocity scale, which sets
the velocities of the nearby \htwo\ Lyman (7-0) P(2) $\lambda$1016.46 and P(3)
$\lambda$1019.50 lines.  The (6-0) P(3) $\lambda$1031.19 and R(4)
$\lambda$1032.35 lines were then used to establish the \OVI\ velocity, as shown
in Fig.  2.  The measured equivalent widths of several interstellar metal lines
are also given in Table 1.

The adopted column densities for several ionic species, as well as several
rotational states of \htwo\, along this sightline, are given in Table 2.  These
were calculated by fitting to a single-component Doppler-broadened curve of
growth for \FeII\ and H$_2$, and by using the apparent column density method
(Savage \& Sembach 1991) for the other species listed.  The two values given for
the \OVI\ column are based on the high and low continuum placements.  We have
added a systematic error of 0.04 dex in quadrature with the statistical error
for \OVI\ to account for errors in continuum placement and the velocity interval
over which the apparent column density is integrated.

\section{Discussion}

Since 114 eV are required to convert O$^{+4}$ to O$^{+5}$, \OVI\ is almost never
produced by photoionization from starlight.  Thus, it is a sensitive tracer of
hot ($\sim300,000$ K) collisionally ionized gas in the interstellar medium.
Adopting the high continuum placement shown in Fig.  2, which we believe is more
appropriate, the \OVI\ column density is $\log N(\mbox{\OVI}) = 14.40\pm0.02$
and $13.89\pm0.03$ in the Milky Way and LMC, respectively.  For the MW gas,
$\log [N(\mbox{\OVI})$sin$|b|]$ = 14.17, which agrees well with the median value
of 14.21 along 11 Galactic halo sightlines studied by Savage et al.  (2000).
Here $b$ = -36\fdg32 is the galactic latitude.

Using $\log N(\mbox{\CIV}) = 14.41$ in the Milky Way from Wakker et al.  (1998)
and assuming an error of 0.05 dex, we find $N(\mbox{\CIV})/N(\mbox{\OVI}) =
1.00\pm0.16$.  If the low continuum is adopted, this ratio is $1.23\pm0.20$.
Either value is greater than, but consistent with, the halo value $\langle
N(\mbox{\CIV})/N(\mbox{\OVI}) \rangle \sim 0.6$ determined from \fuse\
observations of four extragalactic sightlines (Savage et al.  2000), as well as
$\langle N(\mbox{\CIV})/N(\mbox{\OVI}) \rangle \sim 0.9$ from \copernicus\
and \iue\ observations of stars in the lower halo (Spitzer, 1996).

Note that the widths of \OVI\ components in both the MW and LMC are much
broader than the thermal line width, which is $\approx$30 \kms\ (FWHM) for gas
at 300,000 K.  The substantial difference in the widths of the \CIV\ and \OVI\
lines may be due to the different scale heights in the Galactic halo of these
ions (Savage et al.  2000).

Sk~-67~05 was the only star for which Wakker et al.  (1998) did not detect
\CIV\ in their study of the LMC halo.  We recalculated their upper limit to
$N(\mbox{\CIV})$ assuming the velocity range observed in the \OVI\ gas ($+180$
to $+322$ \kms) and find $\log N(\mbox{\CIV}) < 13.5 \ (3\sigma)$.  Adopting
the high continuum case we find $N(\mbox{\CIV}) / N(\mbox{\OVI}) < 0.4 \
(3\sigma)$ for the LMC material along the Sk~-67~05 sightline.

The \htwo\ column density varies greatly between the LMC and Milky Way.  For the
LMC we find $\log N({\rm H}_2) = 19.50\pm0.08$, summed over the $J=0$ to 5
states (Table 2), and a $b$-value of $5\pm2$ \kms.  This column density is
significantly higher than that measured along other LMC sightlines (de Boer et
al.  1998; Shull et al.  2000).  By comparison, \fuse\ observations of the metal
deficient galaxy I Zw 18, have yielded only an upper limit, $\log N({\rm H}_2)
\lesssim\ 15$ (Vidal-Madjar et al.  2000).

The distribution of \htwo\ rotational states in the LMC is best fitted by a
two-component excitation temperature (Fig.  4) with $T_{01}=59\pm5$ K for
$J=0,1$ and $T_{ex} = 800\pm330$ K for $J=3,4,5$.  This is similar to the
temperature distribution seen along the sightline to star LH10:3120 in the LMC
using ORFEUS (de Boer et al.  1998).  This indicates that the gas is not in
thermal equilibrium, and the higher $J$ states are fluorescently pumped by UV
radiation.  This excited gas may therefore exist in the outer, optically thin
regions of the cloud.  In the interior regions the molecular fraction may
increase due to self-shielding, which is expected when E(\bv) $\ga$ 0.1 (Savage
et al.  1977).  The $J=2$ point falls almost exactly on the extension of the
$T_{01}$ line, suggesting that the density of this gas is rather high.

For the Milky Way gas we find an H$_2$ column density of $\log N({\rm H}_2) =
14.95\pm0.08$, summed over the $J=0$ to 3 states (Table 2), and a rotational
temperature of $T_{01} = 99^{+30}_{-20}$ K (Fig.  4).  This temperature is
similar to other values measured with \fuse\ (Shull et al.  2000) and to
\copernicus\ measurements of Galactic stars (Savage et al.  1977).  There is an
indication of a two-component temperature distribution for the Milky Way gas as
well.  However, because of the relatively small number of measurable lines and
severe blending, we are unable to accurately determine an excitation temperature
for the higher-$J$ MW material.

Absorption from both Milky Way and LMC material is clearly seen in all of the
low ions present in the LiF1 data (e.g., Figs.  1 and 3).  Blending of atomic,
ionic, and \htwo\ absorption along this sightline limits the number of species
for which we can derive accurate equivalent widths and column densities.  The
depleted species \FeII\ has several well-observed transitions in our dataset
(Table 1).  We have constructed a single-component Doppler-broadened curve of
growth for the \FeII\ lines observed in both the Milky Way and the LMC.  The
best fit yields $\log N(\mbox{\ion{Fe}{2}}) = 14.8\pm0.1$ and $14.8\pm0.1$ with
$b=14.2\pm0.5$ and $12.1\err{2.1}{1.6}$ \kms\ for the Milky Way and LMC,
respectively.  Due to potential uncertainties in some of the $f$-values we have
adopted a conservative error of 0.1 dex in these column densities.  Our data
permit good measurements of the non-depleted species \ion{P}{2} $\lambda$1152.82,
and we derive lower limits to the Milky Way and LMC column densities (Table 2)
using the apparent column density method of Savage \& Sembach (1991).  If the
$b$-values for \ion{P}{2} are similar to those derived for \FeII, the data
suggest only very moderate saturation corrections of $\la0.1$ dex.

For the sightline through the halo of the Milky Way, we derive a gas-phase
abundance ${\rm [Fe/P]} \equiv \log N(\mbox{\ion{Fe}{2}}) - \log
N(\mbox{\ion{P}{2}}) - \log ({\rm Fe/P})_\odot \sim -0.6$, assuming a solar
system ratio of $\log ({\rm Fe/P})_\odot = +1.92$ (Anders \& Grevesse 1989).
For the LMC gas along this direction we find ${\rm [Fe/P]} \sim -0.7$.  This
suggests significant incorporation of iron into dust grains in both galaxies.
Differences in the relative abundance of singly- and doubly-ionized iron and
phosphorous could affect this measurement (Sembach et al.  2000), but the
magnitude of this effect is likely to be much too small to account for the
gas-phase deficiency of iron along this sightline.  The derived values of [Fe/P]
are similar to the values of ${\rm [Fe/Zn]} = -1.06$ and $-0.91$ for the Milky
Way halo and LMC absorption towards SN 1987A (Welty et al.  1999).  Future
\fuse\ observations of a large number of LMC sightlines will allow us to study
the distribution of gas-phase abundances and infer the composition of
interstellar dust in this environment.

\section{Summary}

We have reported equivalent widths and column densities of \OVI, \htwo,
\ion{P}{2}, \ion{Si}{2}, \ion{Ar}{1}, and \ion{Fe}{2} along the line of sight to
Sk~-67~05 in the LMC using \fuse\ data.  The principal results of this study
are:

\begin{enumerate} \item In the halo of the Milky Way toward the LMC
$N(\mbox{\CIV})/N(\mbox{OVI}) = 1.00\pm0.16$, a value somewhat greater than but
consistent with other \fuse\ observations through the halo (Savage et al.
2000).  In the LMC, where only an upper limit on $N(\mbox{\CIV})$ is available,
we find $N(\mbox{\CIV}) / N(\mbox{\OVI}) < 0.4 \ (3\sigma)$.  This is
consistent with the lower ratio seen in the disk of the Milky Way compared to
the halo (Savage et al.  2000; Spitzer 1996).

\item The LMC is rich in \htwo\ along this sightline, $\log N({\rm H}_2) =
19.50\pm0.08$. A two-component temperature fit gives $T_{01}=59\pm5$ K for and
$T_{ex} = 800\pm330$ K.

\item The gas-phase abundances of ${\rm [Fe/P]} \sim -0.6$ and $\sim -0.7$ in
the Milky Way and LMC suggests significant depletion in both locations relative
to solar abundances.  \end{enumerate}

\acknowledgements

We thank D.  Massa for providing the software to co-add the spectra used in
this analysis.  This work is based on data obtained for the Guaranteed Time Team
by the NASA-CNES-CSA \fuse\ mission operated by the Johns Hopkins University.
Financial support to U.  S.  participants has been provided by NASA contract
NAS5-32985


\begin{deluxetable}{lcccc}
\tablenum{1}
\tablecolumns{5}
\tablewidth{0pc}
\tablecaption{Ionic Equivalent Widths}
\tablehead{
\colhead{} & \colhead{} & 
\colhead{} & 
\multicolumn{2}{c}{$W_\lambda$ [m\AA]\tablenotemark{a}} \\
\cline{4-5}
\colhead{Ion} & \colhead{$\lambda$} & \colhead{$\log \lambda f$} & 
\colhead{MW} &
\colhead{LMC}
}
\startdata
\ion{O}{6} & 1031.926 & 2.14
	& $249\pm10$\tablenotemark{b} & $87\pm5$\tablenotemark{b} \\
	& 1031.926 & 2.14
	& $213\pm8$\tablenotemark{c}  & \nodata \\
\ion{Si}{2} & 1020.699 & 1.46 & $99\pm3$ & $143.0\pm2.4$ \\
\ion{P}{2}  & 1152.818 & 2.43 & $74.3\pm2.1$ & $78.6\pm1.7$ \\
\ion{Ar}{1} & 1048.220 & 2.41 & $127.0\pm2.2$ & $105.2\pm2.4$ \\
\ion{Fe}{2} & 1055.262 & 0.81
		& \nodata & $31.0\pm1.9$ \\
\ion{Fe}{2} & 1112.048 & 0.84\tablenotemark{d}
		& \nodata & $34.5\pm1.9$ \\
\ion{Fe}{2} & 1121.975 & 1.35\tablenotemark{d} 
		& $85.4\pm1.9$ & $81\pm3$ \\
\ion{Fe}{2} & 1125.448 & 1.24
		& $77.8\pm2.3$ & $66.7\pm2.3$ \\
\ion{Fe}{2} & 1127.098 & 0.48\tablenotemark{d} 
		& $16.2\pm1.9$ & $11.2\pm1.1$ \\
\ion{Fe}{2} & 1133.665 & 0.74 & $38\pm5$     & \nodata \\
\ion{Fe}{2} & 1142.366 & 0.63 & $16\pm3$     & \nodata \\
\ion{Fe}{2} & 1144.938 & 2.10 & $173\pm4$    & \nodata \\
\enddata
\tablenotetext{a}{Measured equivalent widths and $1\sigma$ uncertainties 
        (in m\AA) for the Milky Way (MW) and LMC components towards
        Sk-67~05.}
\tablenotetext{b}{Derived using the ``high'' continuum (see text).}
\tablenotetext{c}{Derived using the ``low'' continuum (see text).  
	Probable contamination from H$_2$ ($6-0$) P(3) 1031.19 \AA\
	line removed.}
\tablenotetext{d}{Oscillator strength taken from the preliminary results
	of Howk et al. (2000, in preparation).}
\end{deluxetable}

\begin{deluxetable}{lccc} 
\tablenum{2} 
\tablecolumns{4} 
\tablewidth{0pc}
\tablecaption{Adopted Column Densities} 
\tablehead{ \colhead{} &
\multicolumn{2}{c}{$\log N$ [cm$^{-2}$]\tablenotemark{a}} \\ \cline{2-3}
\colhead{Species} & \colhead{MW} & \colhead{LMC} &
\colhead{Method\tablenotemark{b}} } 
\startdata 
\ion{O}{6} &
$14.40\pm0.02$\tablenotemark{c} & $13.89\pm0.03$\tablenotemark{c} & 1 \\
\ion{O}{6} & $14.32\pm0.02$\tablenotemark{d} & \nodata & 1 \\ 
\ion{P}{2} & $\ga
13.5$ & $\ga 13.6$ & 1 \\ 
\ion{Si}{2} & $\ga 13.9$ & $\ga 14.5$ & 1 \\
\ion{Ar}{1} & $\ga 14.0$ & $\ga 13.8$ & 1 \\ 
\ion{Fe}{2} & $14.8\pm0.1$ & $14.8\pm0.1$ & 2 \\ 
H$_2$ ($J=0$) & $14.26\pm0.09$ & $19.34\pm0.10$ & 2 \\
H$_2$ ($J=1$) & $14.46\pm0.08$ & $18.99\pm0.10$ & 2 \\ 
H$_2$ ($J=2$) &
$14.41^{+0.15}_{-0.21}$ & $16.28^{+0.30}_{-0.50}$ & 2 \\ 
H$_2$ ($J=3$) &
$14.23^{+0.13}_{-0.19}$ & $15.28^{+0.30}_{-0.50}$ & 2 \\ 
H$_2$ ($J=4$) & \nodata
& $14.46\pm0.06$ & 2 \\ H$_2$ ($J=5$) & \nodata & $14.62^{+0.18}_{-0.30}$ & 2 \\
\enddata 
\tablenotetext{a}{Logarithm of the adopted column densities for the
Milky Way (MW) and LMC components towards Sk-67~05 (in units of ions
cm$^{-2}$).}  \tablenotetext{b}{Method used for deriving column densities:  (1)
Apparent column density method; (2) Curve of growth fitting method.}
\tablenotetext{c}{Column density using ``high'' continuum (see text).}
\tablenotetext{d}{Column density using ``low'' continuum (see text).}
\end{deluxetable}


\begin{figure}
\epsscale{0.7}
\plotone{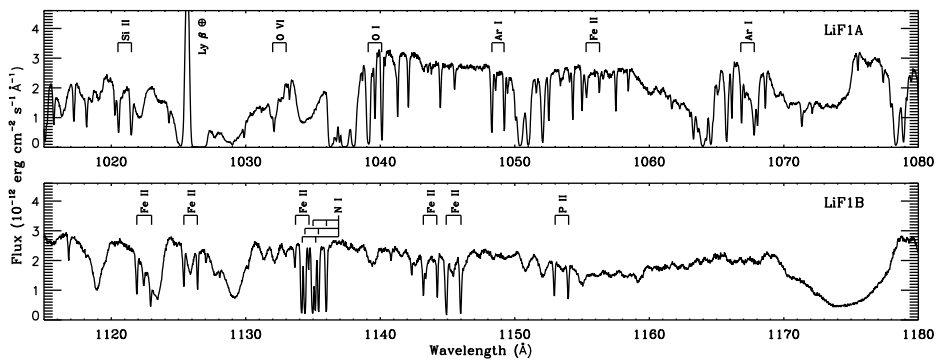}
\caption{The composite LiF1A and LiF1B spectra of Sk~-67~05.  Several 
important metal lines having components in both the Milky Way and LMC
are marked.  Most of the other absorption lines are due to molecular
hydrogen in the LMC.  The strong emission line at1025.72 \AA\ is due
to terrestrial Ly$\beta$ airglow. The fully sampled data has been 
smoothed by four pixels in these plots.  This smoothing does not
affect the observed linewidths because the data are oversampled. 
\label{fig:one}}
\end{figure}

\begin{figure}
\epsscale{0.75}
\plotone{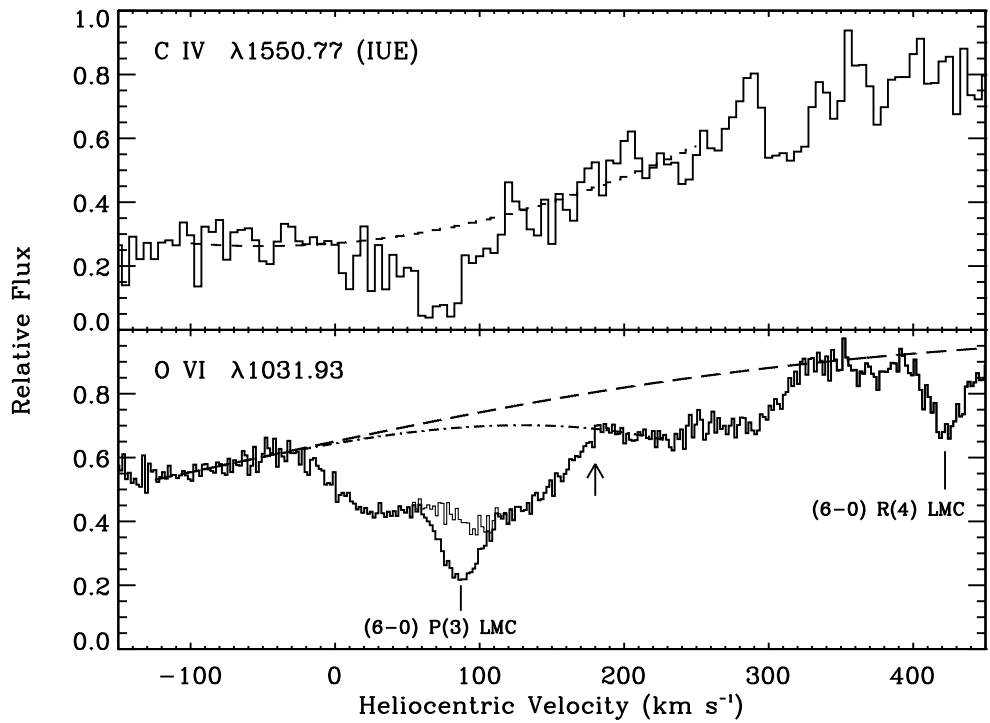}
\caption{The \protect\OVI\ 1031.93 \AA\ line profile from \protect\fuse\ and the
\protect\CIV\ 1550.77 \AA\ line profile from \protect\iue.  The velocity scale
has been adjusted as described in the text.  Both ``high'' (long-dashed line)
and ``low'' (dashed-dotted line) continuum placements have been considered in
the \protect\OVI\ analysis.  The arrow denotes the adopted division between the
Milky Way gas at velocities $<180$ \protect\kms\ and the LMC gas at velocities
$>180$ \kms.  The \protect\htwo\ ($6-0$) P(3) line, arising in the LMC, falls
directly in the Milky Way \protect\OVI\ absorption line.  The model fit has been
divided out to remove the effects of the \htwo\ absorption, and the resulting 
\OVI\ profile is shown as a light line.
\label{fig:two}}
\end{figure}

\begin{figure}
\epsscale{0.8}
\plotone{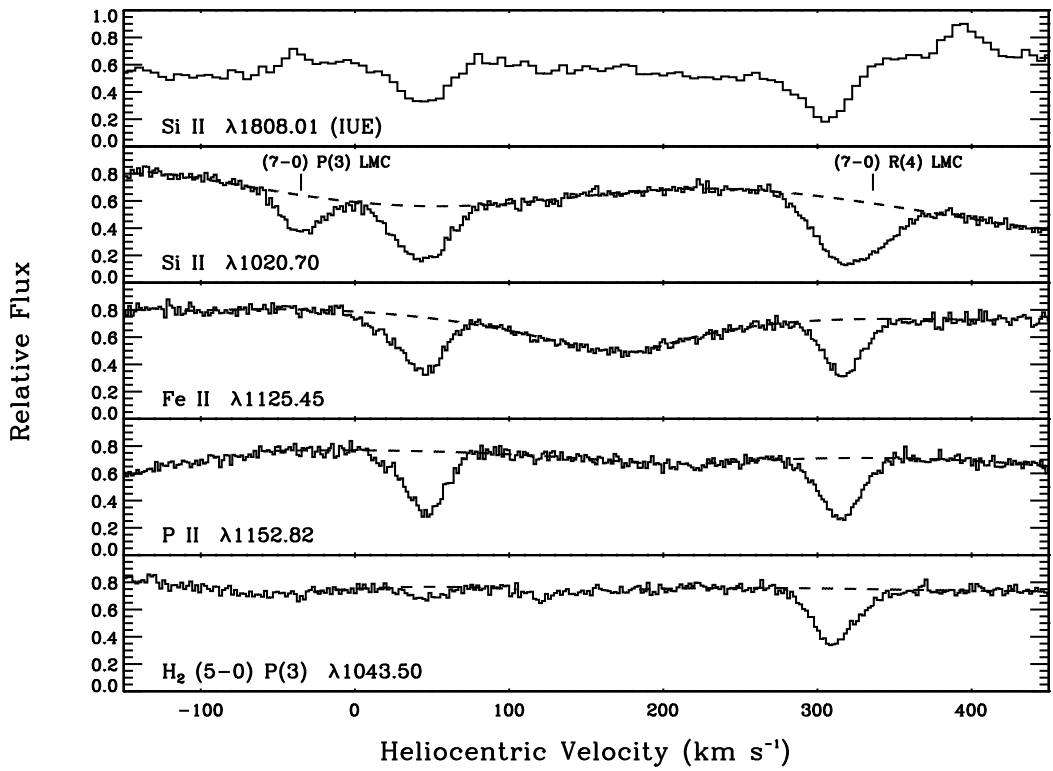}
\caption{Selected absorption lines of low ions in the \protect\fuse\ bandpass,
together with the \protect\ion{Si}{2} 1808.01 \AA\ line from \iue\ observations.
Dashed lines show the adopted continuum levels. The equivalent widths of the MW and LMC components are given in Table 1.  The
velocity scales of the \protect\fuse\ lines have been adjusted to make the Milky
Way component velocities match that of the \protect\ion{Si}{2} \protect\iue\
line.  The LMC component of \protect\ion{Si}{2} $\lambda1020.70$ is broadened
due to a blend with the indicated \htwo\ line.
\label{fig:three}}
\end{figure}

\begin{figure}
\epsscale{0.8}
\plotone{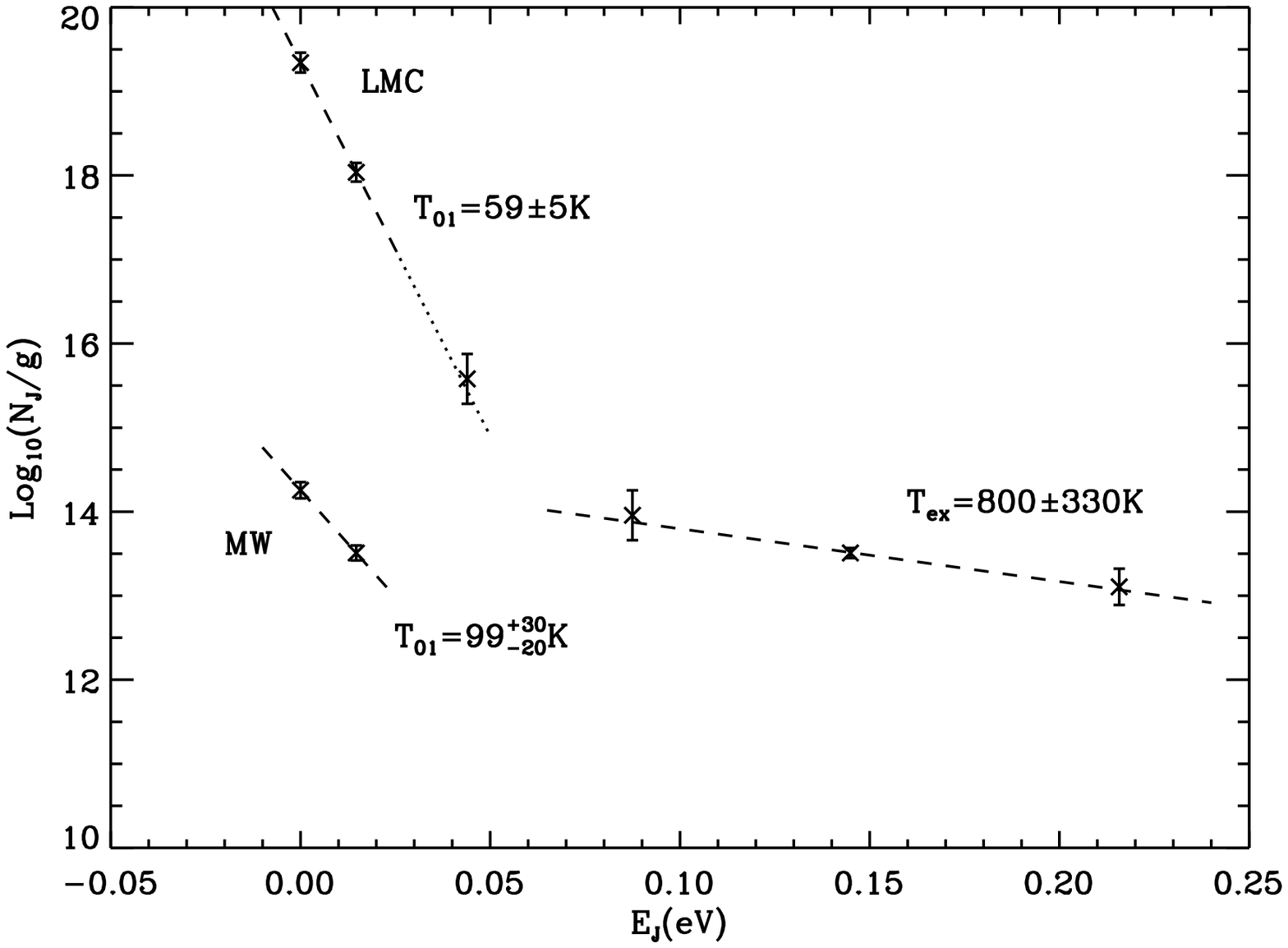}
\caption{Column densities for the $J$=0-5 levels of H$_2$ in the LMC and 
Milky Way.  For the LMC a two-component fit is possible, with the
excitation temperatures indicated.  For the much weaker Milky Way
lines, reliable column densities could only be determined for the $J$=0
and 1 levels.  Note that the dotted line extension of the LMC T$_{01}$
line intersects the $J$=2 point, indicating that the density of this gas
is rather high. The LMC H$_2$ column densities shown here are
significantly higher than those reported along other LMC sightlines
(de Boer et al. 1998; Shull et al. 2000).
\label{fig:four}}
\end{figure}


\begin{references}

\reference{} Anders, E. \& Grevesse, N. 1989, Geochim. Cosmochim. Acta, 53, 197

\reference{} Binney, J. \& Merrifield, M. 1998, {\em Galactic Astronomy}, 
	Princeton University Press.

\reference{} de Boer, K. S., Richter, P., Bomans, D. J., Heithausen, 
	A., \& Koorneef, J. 1998, \aap, 338, L5

\reference{} Chu, Y.-H., Wakker, B., Mac Low, M.M., Garcia-Segura, G. 
	1994, \aj, 108, 1696
	
\reference{} Diplas, A. \& Savage, B. D. 1994, \apj, 427, 274.

\reference{} Massa, D. et al. 2000, \apj, this issue

\reference{} Moos, H. W. et al. 2000, \apj, this issue

\reference{} Sahnow, D. J. et al. 2000, \apj, this issue

\reference{} Savage, B. D., Bohlin, R. C., Drake, J. F., \& Budich, W. 1977,
\apj, 216, 291

\reference{} Savage, B. D. \& de Boer, K. S. 1979, \apj, 230, L77

\reference{} Savage, B. D.\& de Boer, K. S. 1981, \apj, 243, 460

\reference{} Savage, B. D. \& Sembach, K. R. 1991, \apj, 379, 245

\reference{} Savage, B. D. et al. 2000, \apj, this issue

\reference{} Sembach, K. R., Howk, J. C., Ryans, R. S., \& Keenan, F. P. 2000,
\apj, 528, 310

\reference{} Shull, J. M. et al. 2000, \apj, this issue

\reference{} Smith Neubig, M. M.,  \& Bruhweiler, F. C. 1999, \aj, 117, 2856

\reference{} Snowden, S. L., \& Petre, R. 1994, \apj, 436, 123

\reference{} Spitzer, L. 1996, \apj, 458, L29

\reference{} Vidal-Madjar, A. et al. 2000, \apj, this issue

\reference{} Wakker, B. P., Howk, J. C., Chu, Y.-H., Bomans, D., \& Points, 
S. D. 1998, \apj, 499, L87

\reference{} Welty, D. E., Frisch, P. C., Sonneborn, G., \& York, D. G. 
1999, \apj, 512, 636

\end{references}
\end{document}